\documentclass[twocolumn,pre,superscriptaddress,epfs,aps,showpacs]{revtex4}
\usepackage{amsmath}
\usepackage{graphicx}
\usepackage{color}
\usepackage{epsfig}

\begin{document}

\title{Depletion forces on circular and elliptical obstacles induced by active matter}

\author{L. R. Leite}
\affiliation{Universidade da Integra\c{c}\~{a}o Internacional da Lusofonia Afro-Brasileira, Campus dos Palmares, 
62785-000 Acarape, Cear\'{a}, Brazil}
\affiliation{Universidade Federal do Cear\'a, Departamento de
F\'{\i}sica Caixa Postal 6030, 60455-760 Fortaleza, Cear\'a, Brazil}

\author{D. Lucena}
\affiliation{Universidade Federal do Cear\'a, Departamento de
F\'{\i}sica Caixa Postal 6030, 60455-760 Fortaleza, Cear\'a, Brazil}

\author{F. Q. Potiguar}
\affiliation{Universidade Federal do Par\'a, Faculdade de
F\'{\i}sica, ICEN, Av. Augusto Correia, 1, Guam\'a, 66075-110, Bel\'em, Par\'a, Brazil}

\author{W. P. Ferreira}
\affiliation{Universidade Federal do Cear\'a, Departamento de
F\'{\i}sica Caixa Postal 6030, 60455-760 Fortaleza, Cear\'a, Brazil}

\date{ \today }

\begin{abstract}
Depletion forces exerted by self-propelled particles on circular and elliptical passive objects are studied using numerical simulations. We show that a bath of active particles can induce repulsive and attractive forces which are sensitive to the shape and orientation of the passive objects (either horizontal or vertical ellipses). The resultant force on the passive objects due to the active particles is studied as a function of the shape and orientation of the passive objects, magnitude of the angular noise, distance between the passive objects. By increasing the distance between obstacles the magnitude of the repulsive depletion force increases, as long as such a distance is less than one active particle diameter. For longer distances, the magnitude of the force always decrease with increasing distance. We also found that attractive forces may arise for vertical ellipses at high enough area fraction.
 
\end{abstract}


\maketitle


\section{Introduction}
\label{sec:introduction}

Active matter or self-propelled particles (SPP) refers to systems in which its entities convert internal energy into motion, being therefore out of equilibrium \cite{ramaswamy, Marchetti}. Typical examples of these systems are biological systems \cite{Cates}, where microorganisms such as bacteria and eukaryote cells propel themselves with hair-like structures known as flagella. These microorganisms exhibit a variety of structural and dynamical patterns \cite{SPP9,SPP13,SPP14,SPP16,SPP17,SPP18,SPP19,SPP20}. Recent studies showed that colloidal particles can also behave as self-propelled particles \cite{SPP1, Volpe}. Active colloids can be induced using photoactive materials \cite{SPP2}. Partially coated colloidal particles with platinum and dispersed in $H_{2}O_{2}$ solution are often used as model of self-propelled colloids\cite{SPP5}. The motion of an artificial micro-swimmer that uses chemical reaction catallyzed on its own surface to achieve autonomous propulsion is also experimentally possible\cite{SPP10}. Erbe \textit{et al.} showed \cite{SPP11} that it is possible to induce active colloids with three different driving mechanisms: by gravity, by a gradient in a magnetic field, and by a local chemical reaction. Catalytically active Janus micro-spheres are capable of autonomous motion and can potentially act as carriers for transportation \cite{SPP12}.

When hard colloidal particles are held in suspension, in a bath of smaller passive colloids under Brownian motion, attractive interactions between the large colloidal particles arises by means of depletion forces. This attraction emerges when particles are close to each other. The overlap of the excluded volume around the large particles increases the volume available to the small ones. Hence, the total entropy increases, as shown in the pioneering paper of Asakura and Oosawa \cite{asakuraoosawa} almost fifty years ago. However, the excluded volume effect observed in equilibrium systems can not be the sole cause for the effective interaction between passive colloids in a bath of active particles \cite{angelani2011}. Instead, this effect, added to the peculiar non-equilibrium features of the dynamics of the self-propelled particles, generates an effective attractive or repulsive interaction which depends, \textit{e.g.} on the shape of the passive particles, magnitude of the velocity of the active particles, ratio size between the passive colloids and active particles, and density \cite{Ni,harder2014,Ray}. In spite of its distinct nature, we follow the literature and still name here the effective interaction between passive objects in a bath of active particles as depletion interaction, and also name the force on the passive objects due to the active particles as depletion force. It is worth to mention that the shape of passive objects is indeed a relevant feature that induce modifications in the depletion interaction, as demonstrated very recently on a experimental study of coloidal particles in a passive bath \cite{gratale2016}.

In this work, we analyze the effects of the shape and relative orientation of passive elliptical colloids (PEC) in a bath of active particles in order to determine the effective interaction between such passive particles. We systematically study how the depletion interaction between two passive elliptical colloids behaves in a 2D (two-dimensional) bath of active particles. In most of the cases we find that the depletion interaction is repulsive, even for very short distances between the passive objects. This is very surprising, and opposite to what has been recently observed for the interaction between paralell hard-walls \cite{Ni}, also in a bath of SPP. As will be discussed along the manuscript, the clustering of SPP on the surface of passive objects is the reason for the depletion interaction in a bath of SPP \cite{Potiguar}. The clustering of SPP was well described by an athermal model system \cite{Marchetti}, the same considered in the present work. 

In Sect. \ref{sect:model} we introduce the model system and define the key quantities used to characterize the system. In Sect. \ref{sect:Results}, we present and discuss the numerical results. Our conclusions are given in Sect. \ref{sect:conclusions}. 
						   

\section{Model System}
\label{sect:model}

\subsection{Details of the numerical simulations}
\begin{figure}[h!]
\includegraphics[width=1.0\linewidth]{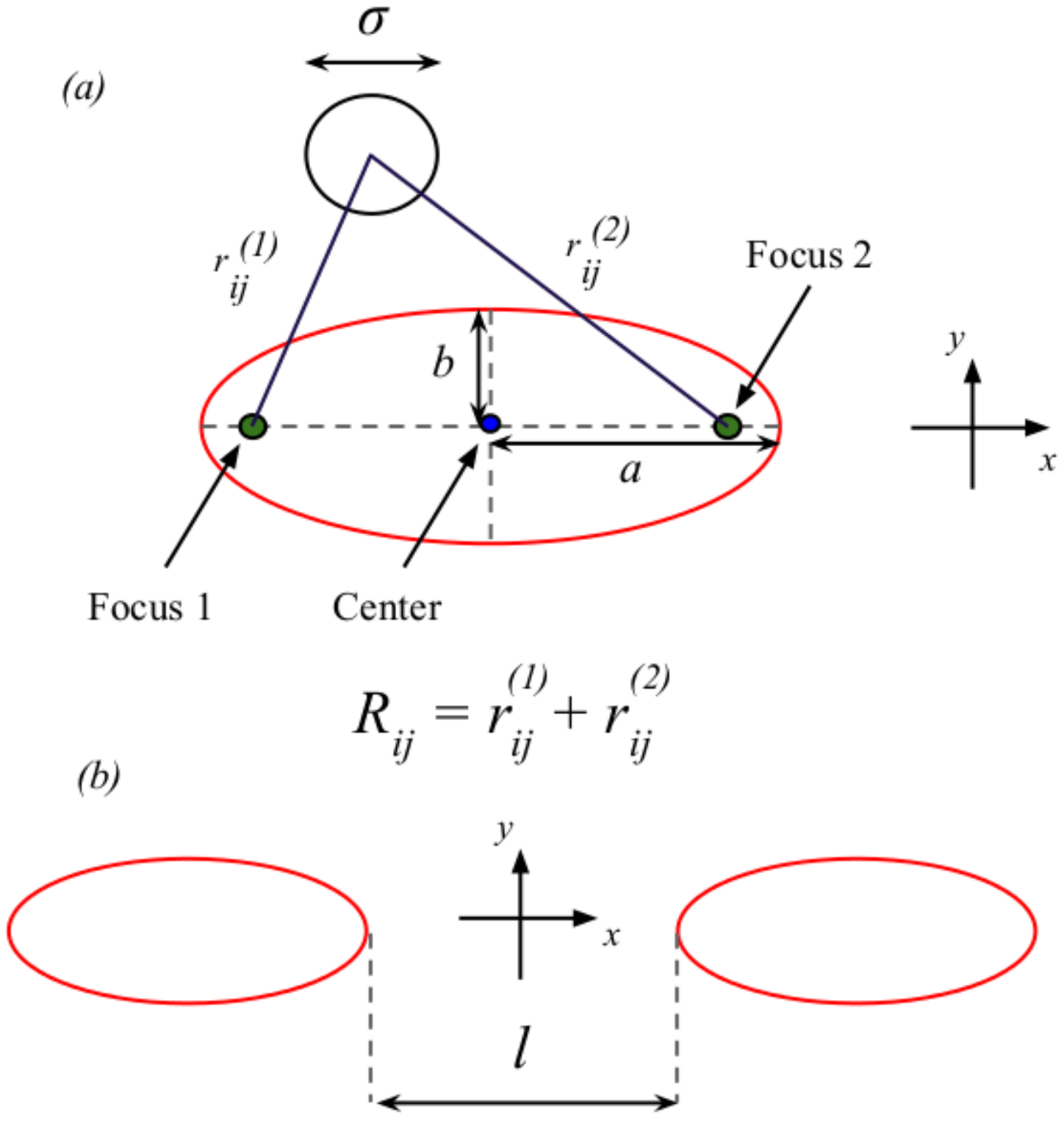}
\caption{(color online) (a) The schematic representation of model system. The black circle represents the SPP of diameter $\sigma$ and red ellipse represents the obstacle. $r^{(1)}_{ij}$ ($r^{(2)}_{ij}$) is the distance between the SPP and the Focus 1 (Focus 2) of the obstacle. $a$ is the size of the horizontal semi-axis and $b$ is the size of the vertical semi-axis. (b) indication of the distance $l$ between the closest points of the passive elliptical colloids.}
\label{fig:model}
\end{figure}

Our system consists of two passive elliptical colloids (PEC) in a 2D bath of $N = 1270$ active particles (or SPP). The SPP are modelled as soft disks of diameter $\sigma$ which interact with each other and with the PEC through spring-like forces of stiffness $\kappa_{\text{SPP}}$ and $\kappa_{\text{PEC}}$, respectively, with $\kappa_{\text{PEC}} \gg \kappa_{\text{SPP}}$. Each SPP moves with velocity $\vec{v}_{i} = v_{0}\cos\theta_{i}(t)\hat{i} + v_{0}\sin\theta_{i}(t)\hat{j}$, where $v_{0}$ is the magnitude of the self-propulsion velocity. In every timestep $\Delta t$, the velocity $\vec{v}_{i}$ is subjected to random fluctuations in the direction $\theta_{i}(t)$ which is proportional to a Gaussian white noise ($G_{i}(t)$) satisfying $\langle G_{i}(t)\rangle = 0$ and $\langle G_{i}(t)G_{j}(t')\rangle = 2\eta\delta_{ij}\delta(t - t')$, where $\eta$ is the angular noise intensity. Similarly, $\langle\xi_{i}(t)\rangle = 0$ and $\langle\xi_{i}(t)\xi_{j}(t')\rangle = 2\xi\delta_{ij}\delta(t - t')$, where $\xi$ is the translational noise intensity. Therefore the equations of motion for the $i$th active particle are written as

\begin{equation}
\frac{\partial \vec{r}_{i}}{\partial t} = \vec{v}_{i} + \mu\vec{F}_{i} + \xi_{i}(t)\mbox{,}\quad\quad\quad\frac{\partial\theta_i}{\partial t} = G_i(t),
\end{equation}\label{Eq:Motion}

\noindent where $\mu$ is the SPP motility, $\vec{F}_{i} = \sum_{j}\vec{F}_{ij}$ is the total force on particle $i$ and the sum is over $j \neq i$ SPP and/or PEC, where ${\bf \hat{r}}_{ij}$ is the unitary vector with the direction pointing from the contact point at the surface of the PEC to the center of the SPP, i.e., from $j$ to $i$. The force on the PEC is the negative of this force, i.e., ${\bf F}_{ij}=-\kappa\alpha_{ij}{\bf \hat{r}}_{ij}$, where $\alpha_{ij}$ is the SPP-PEC overlap. When $\alpha_{ij}>0$ then (${\bf F}_{ij}=0$ otherwise). For interaction between SPP $\alpha_{ij}=\sigma-r_{ij}$, where $\sigma$ is the diameter of a single SPP. For SPP-PEC interaction $\alpha_{ij} = R_{ij} - 2z$, where $z = a$ (if $a > b$) and $z = b$ (if $b > a$), and $R_{ij} = r^{(1)}_{ij} + r^{(2)}_{ij}$, where $r^{(1)}_{ij}$ ($r^{(2)}_{ij}$) is the distance between the $i$th SPP and the first (second) focus of the $j$th ellipse (Fig. \ref{fig:model}). 

For horizontal (semi-major axis along the $x$-axis) PEC, $r^{(1)}_{ij}$ and $r^{(2)}_{ij}$ equations are given by 
\begin{equation}r^{(1)}_{ij} = \sqrt{\left(x_{i} - \left(x_{j} - \sqrt{\left(a + \frac{\sigma}{2}\right)^{2} - \left(b + \frac{\sigma}{2}\right)^{2}}\right)\right)^{2} + y_{i}^{2}},\end{equation} 
\begin{equation}r^{(2)}_{ij} = \sqrt{\left(x_{i} - \left(x_{j} + \sqrt{\left(a + \frac{\sigma}{2}\right)^{2} - \left(b + \frac{\sigma}{2}\right)^{2}}\right)\right)^{2} + y_{i}^{2}}.\end{equation}

\noindent where $x_{i}$ and $y_{i}$ are the coordinates of the $i$th SPP, and $x_{j}$ is the coordinate of the $j$th PEC. The PEC are assumed to be always along the $x$-axis ($y_{j} = 0$). For vertical (semi-major axis along the $y$-axis) PEC, $r^{(1)}_{ij}$ and $r^{(2)}_{ij}$ are given by
\begin{equation}r^{(1)}_{ij} = \sqrt{x_{i}^{2} + \left(y_{i} + \sqrt{\left(a - \frac{\sigma}{2}\right)^{2} - \left(b + \frac{\sigma}{2}\right)^{2}}\right)^{2}},\end{equation}
\begin{equation}r^{(2)}_{ij} = \sqrt{x_{i}^{2} + \left(y_{i} + \sqrt{\left(a + \frac{\sigma}{2}\right)^{2} - \left(b + \frac{\sigma}{2}\right)^{2}}\right)^{2}}.\end{equation}

In the present work we consider $\xi_{i}(t) = 0$, which means that the SPP are not submitted to translational thermal fluctuations, and the detailed balance is not obeyed in this system. Recently, such a model system has been called an athermal model system\cite{Marchetti,Potiguar}. 

In all simulations we employed periodic boundary conditions in both $x-$ and $y-$directions. The equations of motion are integrated using a second order stochastic Runge-Kutta algorithm \cite{honeycutt92}. Lengths are given in units of $\sigma$, and the unit of force $F_{0}$ is such that $\kappa_{\text{SPP}} = \tilde{\kappa} (\sigma/F_{0})$ where $\tilde{\kappa}$ has units of force per distance. The unit of time is $t_{0} = \sigma/\mu F_{0}$. Henceforth, all quantities are dimensionless, unless stated otherwise. We consider $v_{0} = 1$, $\kappa_{SPP} = 50$, $\kappa_{PEC} = 1000$ and $\mu = 1$. The equations of motion are integrated using a time step $\Delta t = 10^{-3}$. In all simulations, we run $5 \times 10^{6}$ thermalisation time steps and calculated averages from $5 \times 10^{6}$ up to $15 \times 10^{6}$ time steps.

\subsection{Interaction between the passive colloids}
We are interested in the force between PEC in a bath of active particles. We proceed in this way by considering two PEC at fixed separation and by calculating their interaction with the active particles. Specifically, we calculate the average force $\langle F_{x} \rangle$, along the $x$-direction, exerted on the left ellipse (Fig. \ref{fig:model}) by the bath of active particles, which is the same in magnitude to the average force exerted on the right ellipse.
We define $l$ as the distance between the closest points of the two passive elliptical colloids (Fig. \ref{fig:model}). 

It is convenient to define a dimensionless parameter $\lambda = b/a$. For the case $\lambda = 1$, the PEC are circles. For $\lambda > 1$ the ellipses have their major axis along the $y$-direction (vertical ellipses), while for $\lambda < 1$ the ellipses have their major axis along the $x$-direction (horizontal ellipses). We always consider the minor axis of the PEC equal to 5. In this way, for $\lambda \ge 1$, $a=5$ and $5\le b \le 10$. For $\lambda \le 1$, $b=5$ and $5\le a \le 10$. 

The magnitude of the interaction between the PEC is studied as a function of the dimensionless parameter $\lambda$, the distance between the PEC, $l$, the angular noise intensity $\eta$ and the area fraction $\phi$, with the later defined as $\phi = \frac{N\pi}{4(L^2-S_T)}$, where $N$ is the number of SPP, $L$ is the size of the 2D squared simulation box and $S_{T}$ is the total area occupied by the PEC. Note that $\phi$ is related to the density $n$ of SPP, i.e. $\phi = \pi n/4$, where $n$ is the SPP density.

In order to present how the SPP are distributed over the simulation box over time, it is convenient to introduce the reduced area fraction distribution, defined as $log_{10}[\phi_i/\phi_{bulk}]$, where $\phi_i$ is the time average area fraction in the $i$-th sub-box (the total simulation box is split in small sub-boxes) and $\phi_{bulk}$ is the time average area fraction calculated far from the PEC. $\phi_{bulk}$ is the average area fraction of the four sub-box located in the corners of the simulation box, which are far enough from the PEC in order to "feel" their presence.

\section{Results and Discussions}
\label{sect:Results}

\begin{figure}[t]
\centerline{\includegraphics[scale=0.86]{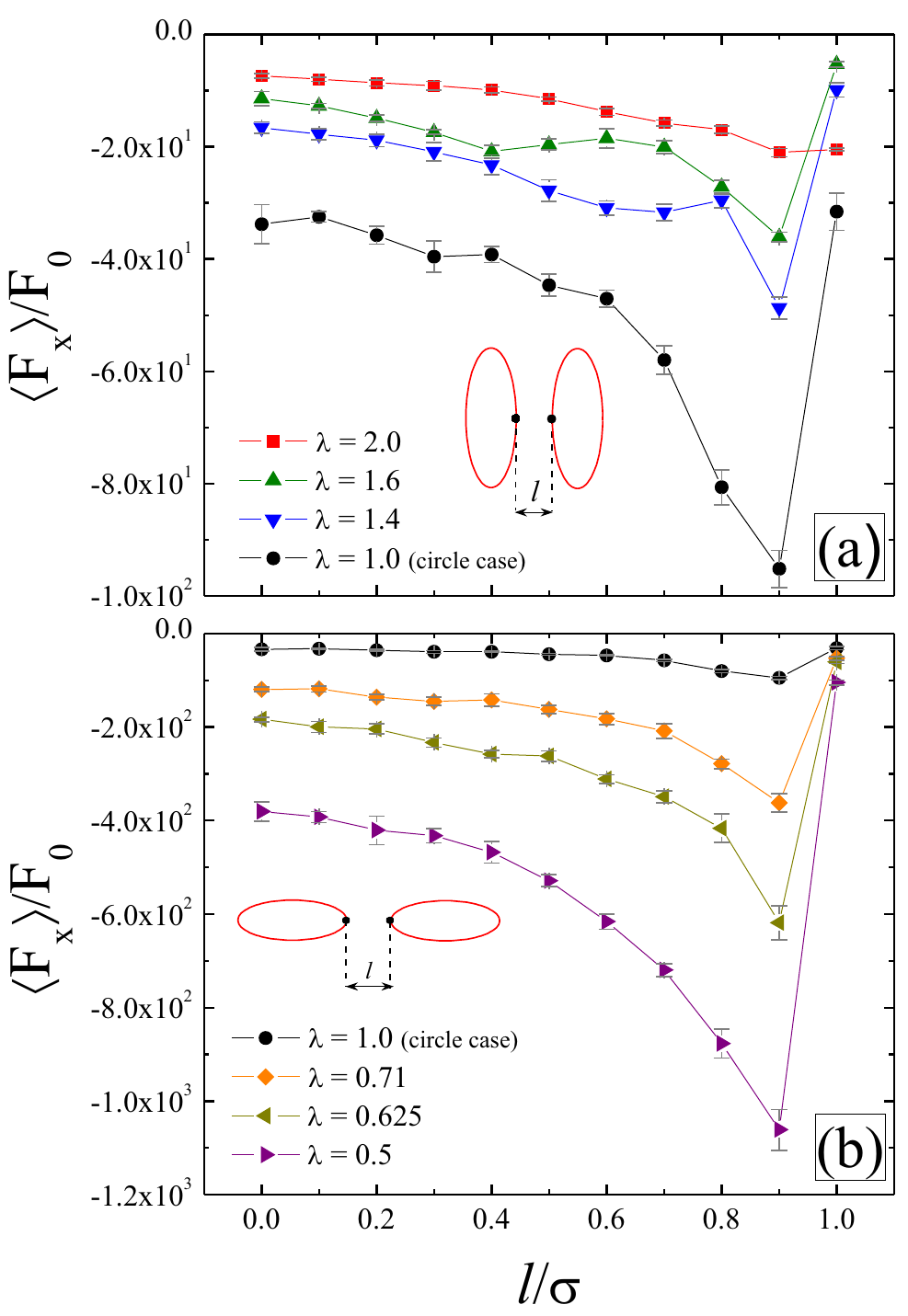}}
\caption{(color online) The average force $\langle F_{x} \rangle$ as a function of the distance between the PEC ($l$), for different values of $\lambda$. (a) Vertical PEC ($\lambda>1$); (b) Horizontal PEC ($\lambda<1$). In both cases, the area fraction of the active bath is $\phi = 0.1$ and the angular noise is $\eta = 10^{-4}$. The force is calculated on the left PEC. Note that Fig.2(a) and Fig.2(b) have different scales, with $\lambda = 1$ curve plotted in both figures.}
\label{fig:fig2}
\end{figure}

\begin{figure}[t]
\centerline{\includegraphics[width=1.0\linewidth]{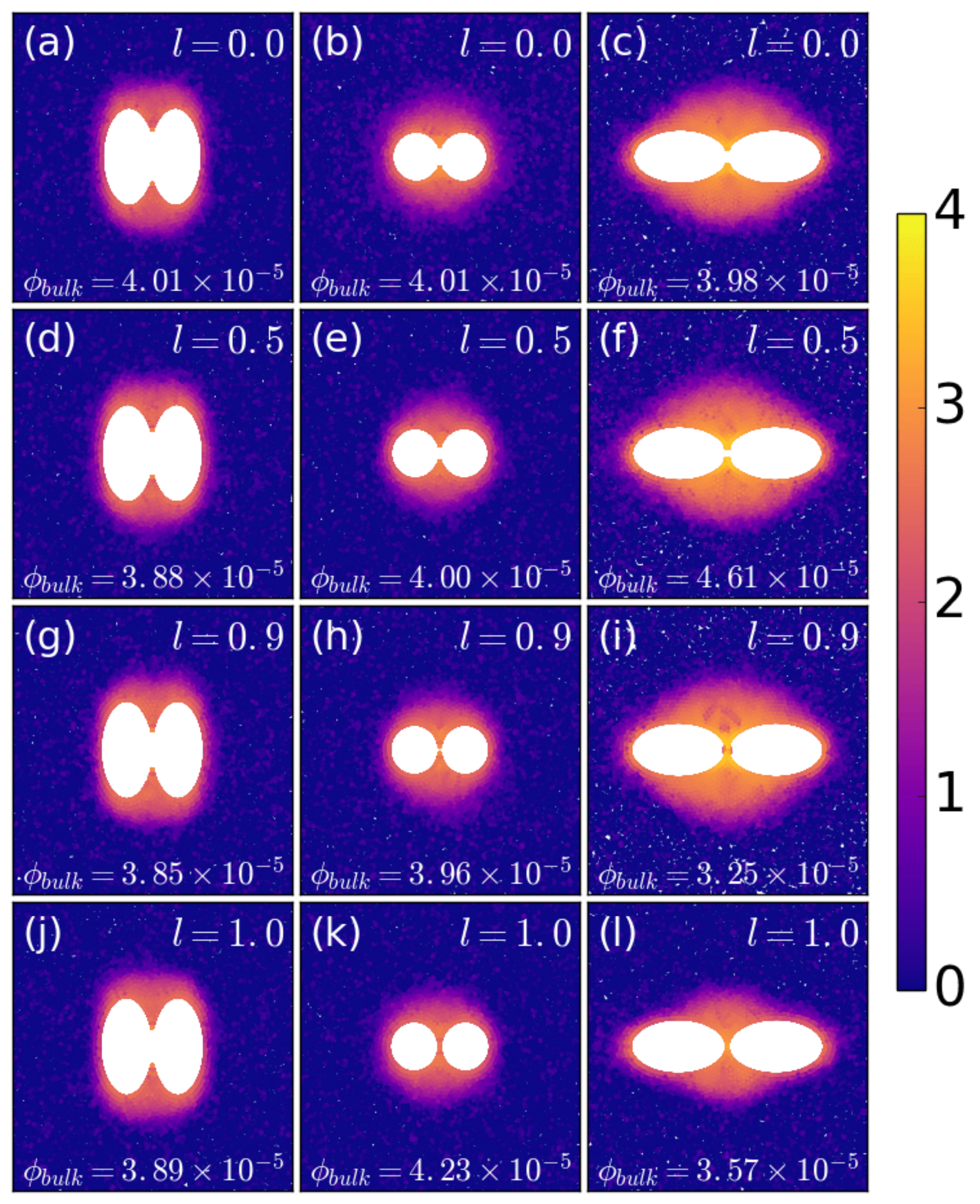}}
\caption{(color online) The reduced area fraction distribution for $l = 0$ and (a)$\lambda = 2$, (b)$\lambda = 1$, (c)$\lambda = 0.5$, $l = 0.5$ and (d)$\lambda = 2$, (e)$\lambda = 1$, (f)$\lambda = 0.5$, $l = 0.9$ and (g)$\lambda = 2$, (h)$\lambda = 1$, (i)$\lambda = 0.5$ and $l = 1$ and (j)$\lambda = 2$, (k)$\lambda = 1$ and (l)$\lambda = 0.5$, for $\phi = 0.1$. We consider a logarithmic plasma color code.} 
\label{fig:fig3}
\end{figure}

\begin{figure}[b]
\centerline{\includegraphics[scale=0.9]{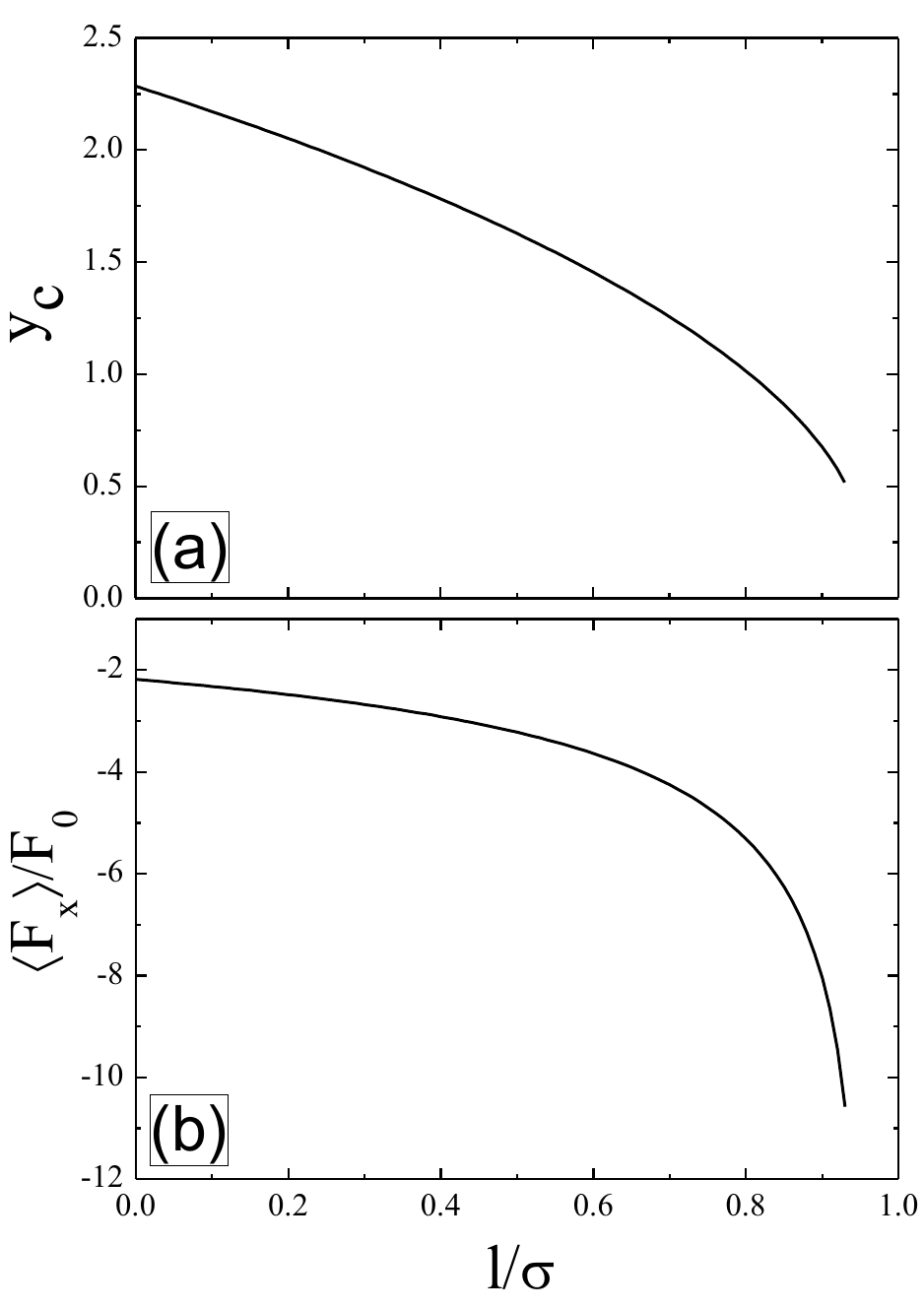}}
\caption{(a) Maximum approach distance, $y_{i}$, as given by the solution of Eq. (6) and (b) depletion force calculated from the compression of a single particle due to the right PEC.}
\label{fig:fig4}
\end{figure}

\subsection{Influence of the shape and orientation of the obstacles}
\label{subsect:shape}

In this section we analyse the influence of the shape and orientation of the PEC on the depletion interaction in the active bath. For this, we study how $\langle F_{x} \rangle$ depends on the separation $l$ (Fig. \ref{fig:model}) between the PEC along the $x$-axis. The shape and orientation of the PEC are controlled by the parameter $\lambda$ which is considered in the interval $0.5 \leq \lambda \leq 2$. The other relevant parameters of the model, namely, the area fraction and noise intensity are kept constant as $\phi = 0.1$ and $\eta = 10^{-4}$, respectively. The average force $\langle F_{x} \rangle$ on the left PEC as a function of $l$ for different values of $\lambda$, is shown in Fig. \ref{fig:fig2}. $\langle F_{x} \rangle$ exhibits negative values for any $\lambda$, indicating that the left PEC is pushed away from the right PEC. The explanation for this behavior consists of two arguments, based on what we observed in our simulations: the first one is based on the aggregation of the SPPs in between the PEC (which is the main reason for the repulsive character of this force); as a consequence of the fact that the SPPs do not collide in the usual sense, but stick to surfaces upon contact. The second argument is stated mainly to explain the general shape of the curves seen in Fig. 2. Therefore, the effective interaction between the PEC is repulsive, which is opposite to the result obtained by Asakura and Oosawa\cite{asakuraoosawa} for a bath of passive particles. Recently, Ni $\textit{et al.}$ showed that the repulsive interaction is not the only one observed in a 2D bath of active particles. In the low density case, it was observed a long-range attraction between parallel hard-wall plates \cite{Ni}. Similar results were also obtained by D. Ray \textit{et al} \cite{Ray} but there they found regimes in which there is a crossover from attraction to repulsion between the walls as a function of wall separation and wall length and by Stenhammar \textit{et al} \cite{Sten}, where it was found that in a mixture of active and passive particles motility of the active component triggers active-passive segregation, which illustrates the attraction between passive particles.

For $l \le 0.9$ the magnitude of $\langle F_{x} \rangle$ increases monotonically with increasing $l$ in both cases $\lambda < 1$ and $\lambda > 1$. $\langle F_{x} \rangle$ drops considerably as the separation between the obstacles approach the diameter of a single SPP. This happens because the SPP becomes able to pass between the obstacles, reducing drastically the pressure on them. It is also interesting to observe that the depletion forces on horizontal PECs are one order of magnitude higher than those on vertical PECs. 

In order to complement and have a better undestanding of the previous results, the reduced area fraction distribution for $l = 0, 0.5, 0.9$ and $l = 1$ is presented in Fig. \ref{fig:fig3}. The concentration of SPP is larger in the region between the PEC compared to the agregation around them. This creates an unbalance on the SPP concentration between PEC such that induces forces on both PEC in opposite directions. In addition, the concentrations of SPP in cases with $\lambda < 1$ are larger than the ones found in cases with $\lambda > 1$. This is one of the reasons for the difference in magnitude of $\langle F_x\rangle$ as a function of $\lambda$ observed in Fig. \ref{fig:fig2}. In addition, the closer the SPP are from the line joining the center of the PEC, the more intense is the {\em x}-component of the depletion force. Figs. \ref{fig:fig3}(a), (d), (g), and (j) indicate that the accumulation of SPP around the vertical PEC is higher in the region around $x = 0$, $y \pm a$, while from Figs. \ref{fig:fig3}(c), (f), (i), and (l) the aggregation of SPP around horizontal PEC is higher around the point $x = 0$, $y = 0$. Therefore, the distribution of SPP around the PEC indicates that the force exerted by the SPP on vertical PEC, is smaller than that exerted on horizontal PEC. 

Regarding the dependence of $\langle F_x\rangle$ on the separation $l$ between the PEC (Fig. 1), we may understand this result qualitatively as follows. When the PEC are at $l = \sigma$, a single particle (we consider it to have very low rotational noise) can fit through them with no overlap, and hence the force is zero. When $l < \sigma$, a particle may still pass between them, but with some overlap (due to the soft-core elastic interaction). Therefore, the repulsive force should increase. This will occur up to a point in which the maximum compression balance the intrinsic force of the SPP, and it gets trapped in between the obstacles. As we lower $l$, the horizontal ($x$-axis) projection of the trapping force decreases, for the SPP will be trapped at a point further away in $y$ from the line joining the PECs; therefore, we should expect a lower repulsion for closer $l$. The active particles get trapped in a similar mechanism as found by Kaiser \textit{et al} \cite{Kaiser}. In our case, vertical PEC, because of the more narrow inner space (compared to that of horizontal PEC), accumulates fewer particles than horizontal PEC, which has a larger inner space.

Moreover remember that we considered very low rotational noise ($\eta = 10^{-4}$) and its variation would only change the magnitude of the force (as detailed in Sect. \ref{subsect:noise}), but not its general dependence on $l$.

Quantitatively, we can frame this discussion by considering a circular PEC ($\lambda = 1.0$), in which an active particle moves in the direction perpendicular to the line joining the obstacles and through the middle point of such line. When it is compressed by the PEC, it experiences a force with a magnitude of $F = \kappa_{\text{PEC}}(\frac{D + \sigma}{2} - r)$, where $D$ is the circular PEC diameter and $r = \sqrt{(D + l)^{2}/4 + y^{2}}$ is the SPP-PEC distance. The vertical force on the active particle due to the compression with the PEC is $F_{y} = Fy/r$, and, in general, we have $F_{y} \le v_{0}\gamma$. When the equality holds, the particle gets trapped in between the PEC. Hence, we can calculate the $y_{i}$ in which this will occur, for a definite $l$, by solving the following equation:

\begin{equation}
v_{0}\gamma = \kappa_{\text{PEC}}\left(\frac{D + \sigma}{2} - r\right)\frac{y}{r}
\label{eq:analytic}
\end{equation}

The repulsive force in each PEC due to this passage of the small particle is simply $F_{x} = F(D + l)/2r$; then, by calculating $y$, and $r$, we can calculate $F_{x}$ . In Fig. \ref{fig:fig4} we present this force and the corresponding solution to Eq. \ref{eq:analytic}, with the parameters used in our simulations. Note that for $l/\sigma > 0.93$, there is no curve because the particle is able to cross the line joining the PEC, resulting that there is no solution to Eq. \ref{eq:analytic}. The curve we obtained has the the features of those seen in Fig. \ref{fig:fig2}, it has a minimum (maximum force) and decreases up to the point in which the PEC touch each other, and the behavior with $l$ is non-linear. The depletion force evaluated above is lower, compared to the numerical results, by, approximately, one order of magnitude. 

These differences between our simulation results and the single particle compression argument we drew are due to the aggregates between the obstacles. This phenomenon will clearly increase the depletion force because more particles will interact with the PEC. 

To make sure the results found in Fig. 2 are induced by the active behaviour of the SPP, we made simulations switching off the self-propulsion with non-zero translational noise (results not shown). The expected atractive interaction between the colloids in a bath of passive particles is indeed observed for $\lambda=1$ and $\lambda= 2$. On the other hand, for $\lambda=0.5$ the depletion force was found to be repulsive, contrary to the well known result of Asakura-Oosawa. The magnitude of this repulsive force is of the order of those we show in Fig. 2(b) ($\langle F_{x}/F_{0}\rangle =-250$ for a thermal noise intersity of 100). We leave the investigation of this result for future work. 

\subsection{Influence of the angular noise $\eta$}
\label{subsect:noise}

Recent studies have shown that the noise intensity plays an important role in active matter systems \cite{Vicsek, Potiguar}. Therefore it is interesting to consider its influence on the depletion interaction in the present model. We study the noise dependence of $\langle F_{x} \rangle$ for $\lambda = 0.5$ (horizontal ellipse), $\lambda = 1.0$ (circle) and $\lambda = 2.0$ (vertical ellipse). The area fraction is $\phi = 0.1$. The PEC are placed in contact with each other, i.e., $l=0$. 

\begin{figure}[t]
\centerline{\includegraphics[width=1.0\linewidth]{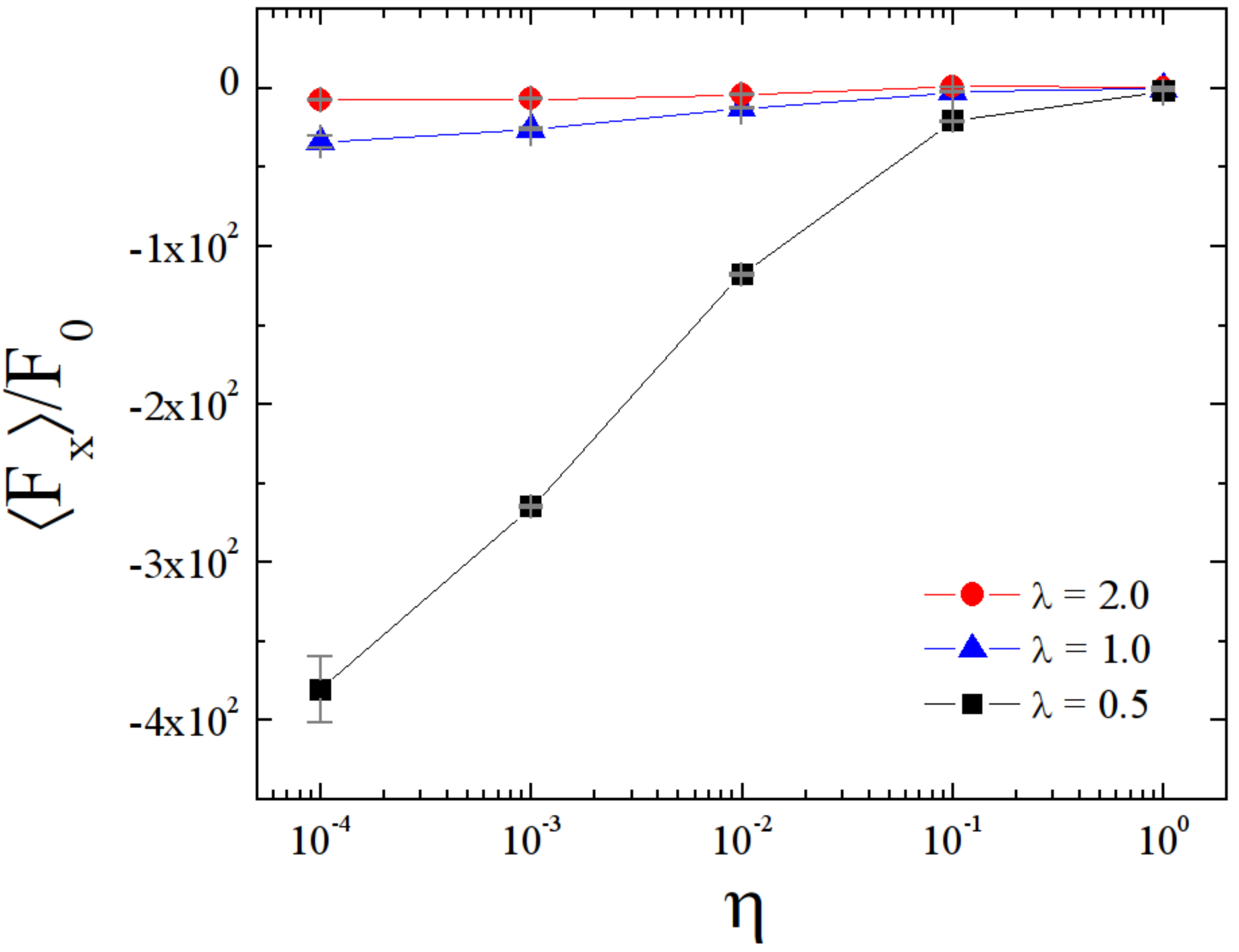}}
\caption{(color online) The average force $\langle F_{x} \rangle$ as a function of the noise, $\eta$ for $\phi = 0.1$. Black squares indicates $\lambda = 0.5$, blue triangles indicates $\lambda = 1.0$ and red circles indicates $\lambda = 2.0$. The obstacles are in contact ($l/\sigma=0.0$).}
\label{fig:fig5}
\end{figure}

The results are presented in Fig. \ref{fig:fig5}. In general, the repulsive force vanishes for a large enough noise intensity. As shown previously \cite{Potiguar}, the dynamics of SSP around rigid obstacles is based on the sliding of the particles over the PEC surface. Large noise intensity results in large fluctuations in the direction of the SPP velocity which allows the SPP to leave the PEC surface, reducing the pressure, and consequently, reducing the repulsive depletion force between the PEC.\\

We also found that with decreasing $\eta$, the magnitude of the force increases much faster for $\lambda=0.5$ than in the other two cases. This is a consequence of the mechanism explained in the Sect. \ref{subsect:shape}. By decreasing the noise intensity large clusters are allowed to form, and these clusters contribute more to the repulsive force in the horizontal case.

\subsection{Influence of the area fraction}
\label{subsect:areafraction}

\begin{figure}[t]
\centerline{\includegraphics[width=1.04\linewidth]{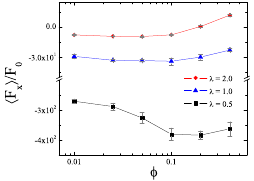}}
\caption{(color online) The average force $\langle F_{x} \rangle$ on the left PEC as a function of the area fraction $\phi$ for $\eta = 10^{-4}$. Black squares indicate $\lambda = 0.5$, blue triangles indicate $\lambda = 1.0$ and red circles indicate $\lambda = 2.0$. The obstacles are in contact ($l/\sigma=0.0$). For $\lambda = 2.0$ the error bars are smaller than the symbols. }
\label{fig:fig6}
\end{figure}

In this section we study the dependence of the average force $\langle F_{x} \rangle$ on the area fraction $\phi$ of the active particles. The results, for the same three $\lambda$ values presented in Sect. \ref{subsect:noise}, are shown in Fig. \ref{fig:fig6} for the case in which the PEC are in contact ($l / \sigma=0$). The angular noise is $\eta =10^{-4}$, which corresponds to the largest magnitude of $\langle F_{x} \rangle$ for both horizontal and vertical PEC (see Fig. 6). 
Beyond the difference of one order of magnitude between the force observed in the cases with vertical and horizontal PEC, the resultant interaction presents a clearqualitative distinction with respect to the shape and orientation of the PEC. For horizontal PEC ($\lambda=0.5$) the depletion force is always repulsive and increases in magnitude with increasing $\phi$. On the other hand, for vertical PEC ($\lambda=2$), the depletion force is repulsive for low $\phi$, and becomes attractive for $\phi \gtrsim 0.2$. A similar qualitative behavior seems to be followed by circular PEC ($\lambda=1$), but we did not found any change from repulsion to attraction in the $\phi$-interval considered in our study ($\phi < 0.4$). Our results are different from the ones found by Ni \textit{et al} \cite{Ni}, in which attractive forces are induced in the low density case in a system of hardwall bars in an active bath. In the present system, we observe only repulsive forces in the low area fraction limit.

\begin{figure}[t]
\centerline{\includegraphics[width=1.0\linewidth]{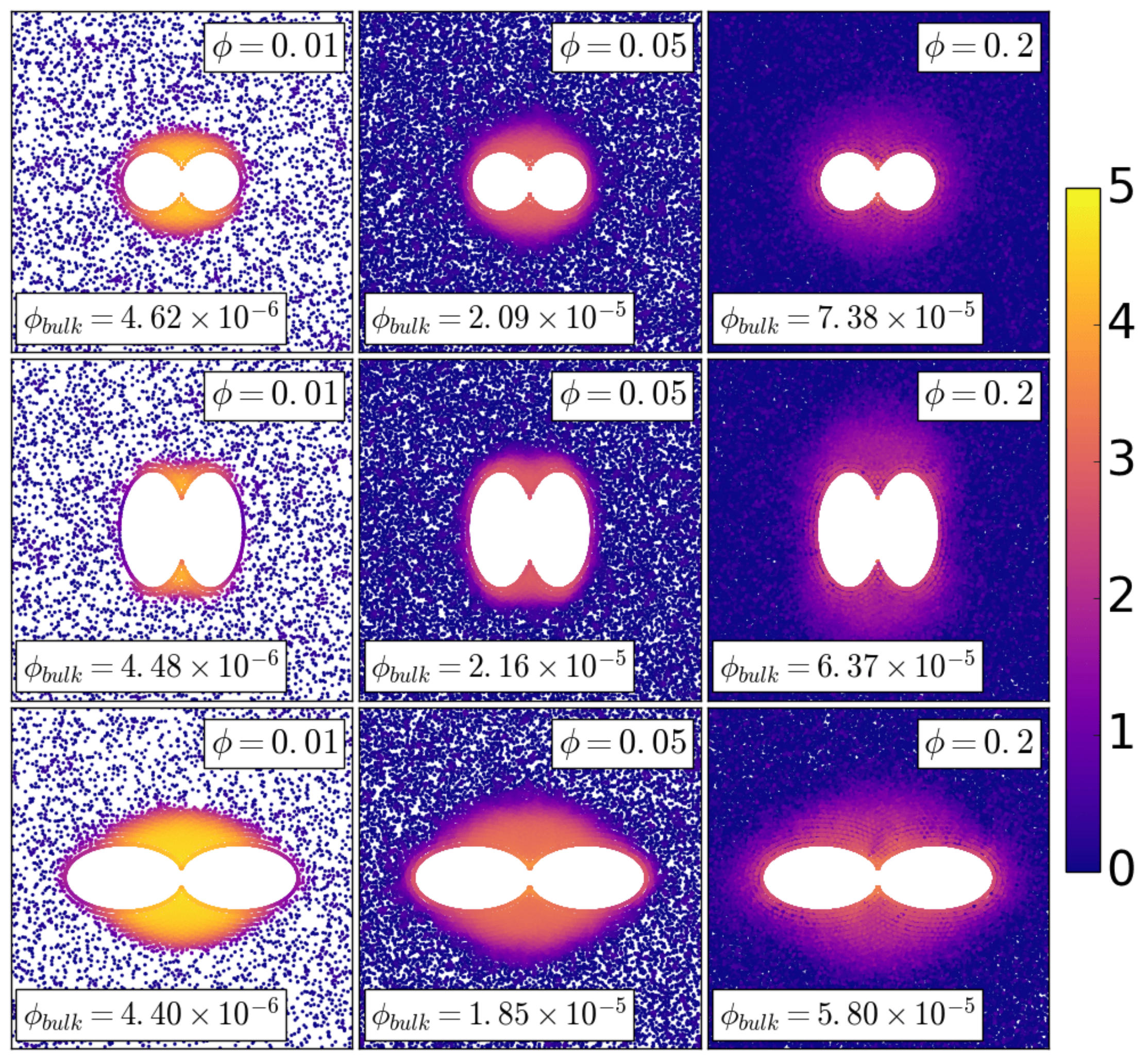}}
\caption{(color online) The reduced area fraction distributions for different area fractions and $\lambda = 1$ (three top panels), $\lambda = 2$ (three middle panels) and $\lambda = 0.5$ (three bottom panels). The angular noise  is $\eta=10^{-4}$. We consider a logarithmic plasma color code.}
\label{fig:fig7}
\end{figure}

Qualitatively, our results can be understood by observing how the SPP are distributed around the PEC. In this case, we resort again on the reduced area fraction distribution, presented in Fig. \ref{fig:fig7} for $\lambda = 2.0$, $\lambda = 1.0$, and $\lambda = 0.5$. The shape and orientation of the PEC is important concerning the accumulation of particles around them. It is remarkable that for any $\phi$ presented in Fig. \ref{fig:fig7} the concentration of SPP around the central region (where the PEC touch each other) is higher in the case of horizontal PEC when compared to the cases of vertical and circular PEC. As a consequence, there is a larger pressure in the central region, explaining the higher repulsive depletion force observed on the horizontal PEC. When the PEC are vertical,  the SPP become more spread around the PEC (and less in the central region) as $\phi$ increases, reducing in this way the pressure in the central region of the PEC and consequently the magnitude of the repulsive force on the PEC. The accumulation of SPP out of the central region between the PEC eventually change the character of the force on the PEC repulsive to attractive. 

\subsection{Depletion forces for $l/\sigma > 1$}
\label{subsect:separation}

\begin{figure}[t]
\centerline{\includegraphics[width=1.0\linewidth]{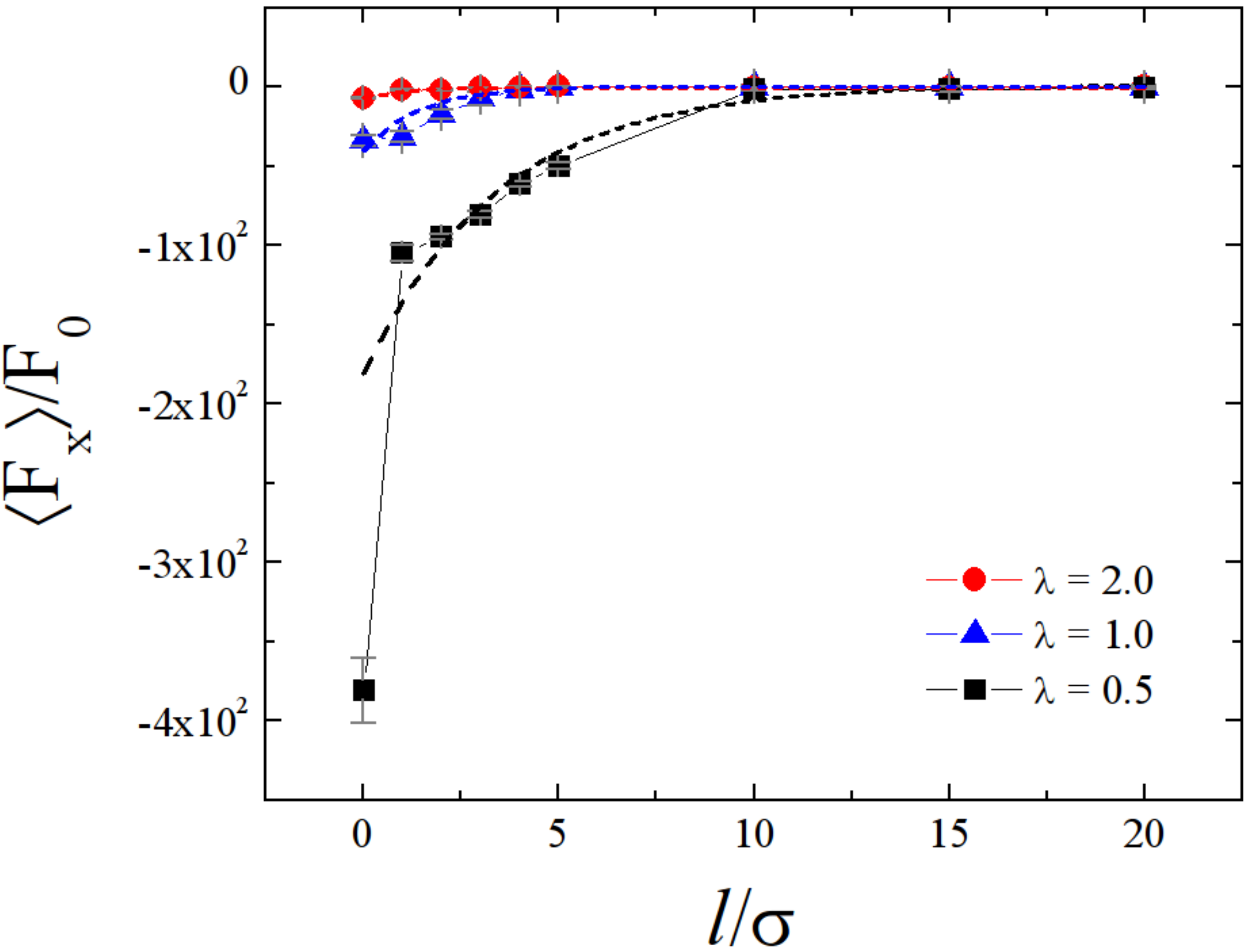}}
\caption{(color online) The average force $\langle F_{x} \rangle$ as a function of the distance $l/\sigma$ for $\phi = 0.1$ and $\eta = 10^{-4}$. Black squares indicates $\lambda = 0.5$, blue triangles indicates $\lambda = 1.0$ and red circles indicates $\lambda = 2.0$. Dashed curves are the exponential fit for $\lambda = 2$ (red) where $\xi \approx 0.89$, $\lambda = 1$ (blue) where $\xi \approx 1.39$ and $\lambda = 0.5$ where $\xi \approx 3.45$ (black).}
\label{fig:fig8}
\end{figure}

\begin{figure}[t]
\centerline{\includegraphics[width=1.0\linewidth]{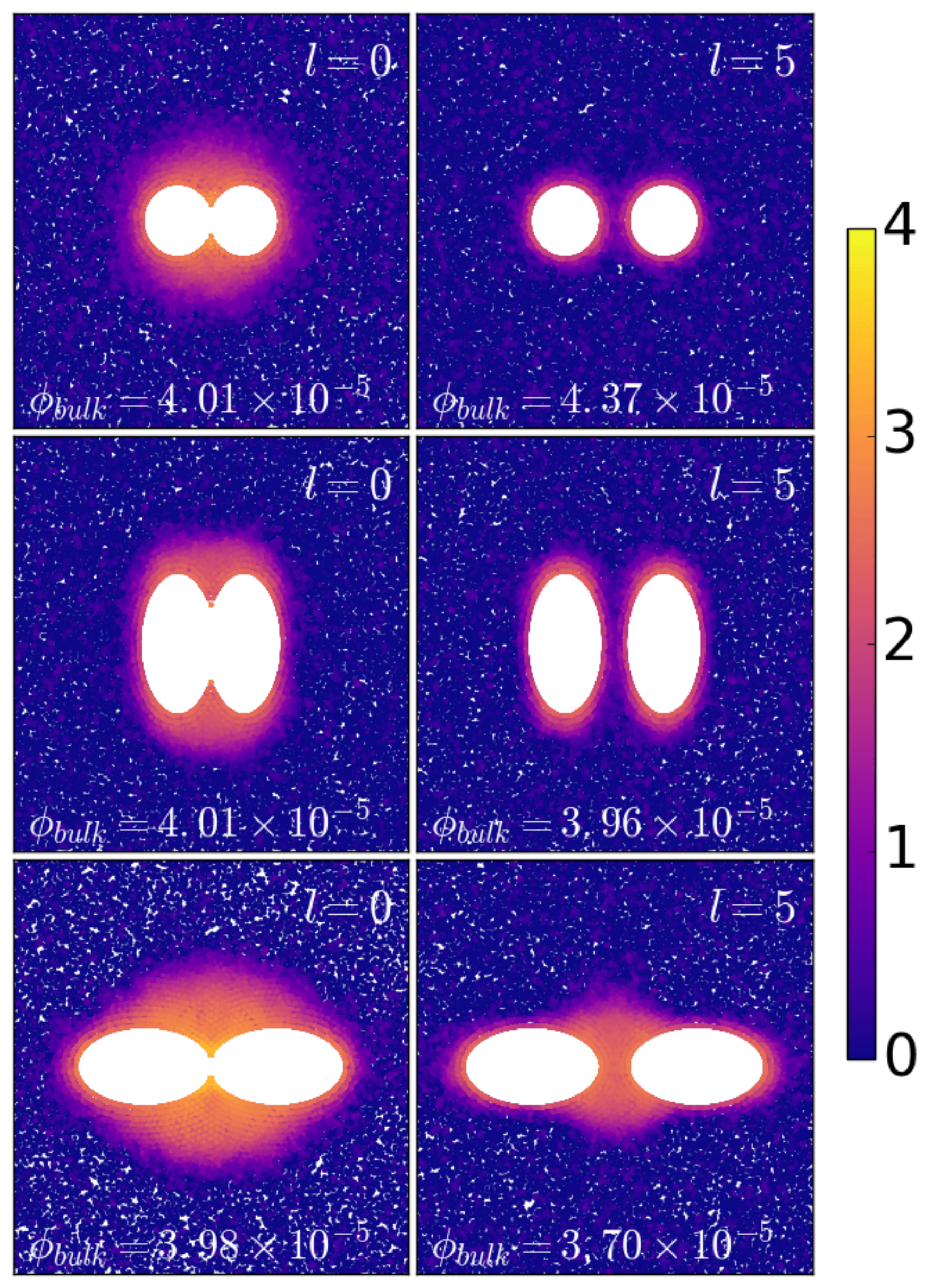}}
\caption{(color online) The reduced area fraction distribution for different distances between the PEC. The area fraction of the SPP is the same for all the presented cases ($\phi = 0.1$). The passive colloids in the top panels have the parameter $\lambda = 1$.  The passive colloids in the middle panels have the parameter $\lambda = 2$, and the passive colloids in the bottom panels have the parameter $\lambda = 0.5$. We consider a logarithmic plasma color code.} 
\label{fig:fig9}
\end{figure}

In this section, the depletion forces are analyzed as a function of separation $l$ between the PEC, but, for $l/\sigma > 1$. The results are shown in Fig. \ref{fig:fig8}.
For circular PEC ($\lambda = 1$) the depletion forces do not vanish up to $l/\sigma = 2$, differently from the analytical prediction by Asakura and Oosawa considering a bath of passive particles.\cite{asakuraoosawa}. Null forces were observed only for larger distances ($l/\sigma \ge 10$ independent of $\lambda$). For the cases shown in Fig. \ref{fig:fig8} $\langle F_{x}\rangle$ decreases, but falls more rapidly for horizontal PEC. For $\lambda=0.5$, $\langle F_{x}\rangle$ takes a longer distance to vanish ($l/\sigma=10$). In Fig. \ref{fig:fig9}, we show the reduced area fraction distribution for $\lambda=2$, $\lambda=1$ and $\lambda=0.5$, and $l/\sigma=0$ and 5. For $\lambda=2$ and $\lambda=1$, each PEC has its own cluster, while for $\lambda=0.5$, there is only one cluster for both PEC, rendering a non-vanishing repulsion at such large distances. It is also important to note that in all three cases as distance $l/\sigma$ increases, depletion forces raises with an exponential behaviour $\langle F_{x}\rangle/F_{0} \propto -\exp^{-l/\xi}$, where $\xi$ can be understood as the range of the force, in agreement with previous works \cite{Ni, harder2014, Ray}. 

\subsection{Depletion forces for vertical-horizontal PECs}
\label{subsect:vertical_horizontal}
\begin{figure}[t]
\centerline{\includegraphics[width=1.0\linewidth]{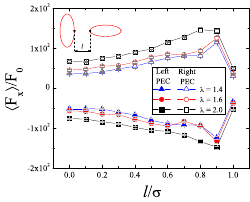}}
\caption{(color online) Average force $\langle F_{x} \rangle$ as a function of the distance $l/\sigma$ for $\phi = 0.1$ and $\eta = 10^{-4}$. $\langle F_{x}\rangle$ on the left PEC is described by full symbols, while for the right PEC, $\langle F_{x}\rangle$ are described by empty symbols. Black squares indicates $\lambda = 0.5$, blue triangles indicates $\lambda = 1.0$ and red circles indicates $\lambda = 2.0$.}
\label{fig:fig10}
\end{figure}

In the previous sections, we showed that forces on horizontal and vertical PECs behave differently as functions of $\lambda$, $\phi$ and $l$. In this section we provide an analysis of the mean depletion forces between PECs, where the left one is vertical and the right one is horizontal. We study the cases where the vertical PEC have $\lambda=1.4, 1.6 , 2.0$ and the horizontal PEC have $\lambda=0.714, 0.625, 0.5$, respectively, such as the product of vertical and horizontal $\lambda$ is always equal to 1. The depletion force as a function of $l/\sigma$ (varying from $0$ to $1$) is shown in Fig. \ref{fig:fig10}.
The results are qualitatively similar to those observed previously when the obstacles are both horizontal or vertical (Fig. \ref{fig:fig2}), i.e, there is a strong increase in magnitude of the repulsive force right below $l=1.0$, and a subsequent decrease of this magnitude as we bring the PECs closer to each other. The main difference between Figs. \ref{fig:fig2} and \ref{fig:fig10} is that the absolute value of the force in Fig. \ref{fig:fig10} is between the values of those curves in Fig. \ref{fig:fig2} for PECs at the same $\lambda$; although they are closer to those values observed for horizontal PEC in Fig. \ref{fig:fig2}. Moreover, we found that the force is the same on both PECs.\\

We should expect that this should occur because none of the ellipses break the symmetry of the system, i.e., there is no induced motion (as would happen if one of those happened to be a half-ellipse). Therefore, the force is equal in both obstacles. Also, the magnitude of this force comes from the aggregate of SPP in between the PECs; now, as we saw in Figs. \ref{fig:fig3}, \ref{fig:fig7}, and \ref{fig:fig9}, this aggregate depends on the shape of the PEC, but as we bring them closer, they merge and form a single structure, exerting the same force in both PECs. From this, we will have an aggregate formed from the cluster of the vertical and the horizontal PEC; and it will be smaller (larger) than between two horizontal (vertical) PEC, and will yield a weaker (stronger) repulsive force.

\section{Conclusions}
\label{sect:conclusions}

We studied depletion forces on passive elliptical colloids (PEC) immersed in a bath of self propelled particles (SPP). In general, the depletion force is repulsive and depends on the shape (eccentricity)  and orientation of the PEC. The dimensionless parameter $\lambda$  (the ratio between the lengths of the semi-major and semi-minor axis of the ellipse) was defined and used to characterize the shape and orientation of the PEC. Horizontal (along the x-axis) and vertical (along the y-axis) PEC are characterized by $\lambda<1$ and $\lambda>1$, respectively. In general, the depletion force observed on horizontal PEC is one order of magnitude larger than that observed on vertical PEC, due to the particular sliding dynamics of active particles in rigid surfaces. We argued that the difference in value of $\langle F_{x}\rangle$ observed for distinct $\lambda$ is due to the larger aggregation (the mechanism responsible for repulsion) in between the horizontal PEC as well as the larger horizontal projection of the compressive force between the SPP and PEC for $\lambda < 1.0$.  With respect to the separation between the PEC, as the PEC get further away from each other, the magnitude of the depletion force increases until the distance reaches $l=0.9$, where the force drops sharply, and almost reaches zero. Such a behavior was observed  independently of $\lambda$. Noise strength has a significant influence on depletion forces, once for the cases studied (circular, vertical and horizontal elliptical objects) the higher the angular noise, the lower the magnitude of the depletion forces. This happens because the clusters around the PEC decrease in size as angular noise raises. By increasing the concentration of active particles (area fraction) we show that it is possible to achieve attractive forces between the PEC for $\lambda \ge 1.0$. High values of the area fraction allow the SPP to coalesce on the outer side of the obstacles, and contribute to lowering the repulsive depletion force. In particular, for $\lambda = 2.0$ (vertical PEC) and $\phi \gtrsim 0.2$, the SPP decrease their presence in the region between the PEC at the same time that they attach to the outer side of the PEC such that the net force becomes attractive. For separation between the PEC $l > 1$, the depletion forces decreases with increasing $l$ but with distinct rates, which depend on the shape of the PEC. E. g., for $l=5$ and $\lambda=0.5$ (horizontal PEC), a considerable agglomeration of SPP is still found between the PEC resulting in a non-vanishing repulsive force, distinctly from the cases $\lambda=1.0$ (circles) and $\lambda=2.0$ (vertical PEC), where $\langle F_{x}\rangle \approx 0$. Finally, we studied the depletion force when one horizontal PEC is close to a vertical PEC in a bath of active particles. The depletion force is repulsive, but with an intermediate magnitude compared to those observed in the cases with two horizontal or two vertical PEC. 

\section{Acknowledgments}
This work was supported by the Brazilian agencies CNPq, CAPES, FUNCAP and FAPESPA. 

\section{Appendix A}

The contact SPP-PEC can be modelled as follows: in order to write the overlap, $\alpha$, as a function of SPP coordinates, we consider that SPP interacts with a new PEC (PEC') such that its semi-axis are $\sigma/2$ bigger than the original one. So $\alpha = r_{ij}^{(1')} + r_{ij}^{(2')} - 2z'^{'}$, where $z' = a'$ for horizontal PEC' and $z' = b'$ for vertical PEC'. Apostrophe (${'}$) means the PEC' variables. Equation for $r_{ij}^{(1')}$  and $r_{ij}^{(2')}$ can be written as
\begin{equation}r_{ij}^{(1')} = \sqrt{(x_{i} - x_{f1}^{'})^{2} + (y_{i} - y_{f1}^{'})^{2}}\end{equation}
\begin{equation}r_{ij}^{(2')} = \sqrt{(x_{i} - x_{f2}^{'})^{2} + (y_{i} - y_{f2}^{'})^{2}}\end{equation}

\noindent where $x_{i}$ and $y_{i}$ are SPP coordinates, $x_{j}$ and $y_{j}$ are PEC coordinates and $x_{f1}^{'}$ and $x_{f2}^{'}$ are PEC' focus coordinates. 

Supposing, at first, a horizontal PEC placed on x-axis, it is important to note that, in this case:\\
(i) ellipse focus coordinates are $(-e'a',0)$ and $(e'a',0)$ (focus 1 and 2, respectively), where $e'$ is the excentricity of PEC'\\
(ii) $e' = \sqrt{a'^{2} - b'^{2}}/a'$, then $x_{f1} = x_{j} - \sqrt{(a'^{2} - b'^{2})}$ and $x_{f2} = x_{j} + \sqrt{(a'^{2} - b'^{2})}$ and\\
(iii) $a^{'} = a + \frac{\sigma}{2}$ and $b^{'} = b + \frac{\sigma}{2}$, so that equations (7) and (8) can be rewritten as
\begin{equation}r^{(1)}_{ij} = \sqrt{\left(x_{i} - \left(x_{j} - \sqrt{\left(a + \frac{\sigma}{2}\right)^{2} - \left(b + \frac{\sigma}{2}\right)^{2}}\right)\right)^{2} + y_{i}^{2}}\end{equation}
\begin{equation}r^{(2)}_{ij} = \sqrt{\left(x_{i} - \left(x_{j} + \sqrt{\left(a + \frac{\sigma}{2}\right)^{2} - \left(b + \frac{\sigma}{2}\right)^{2}}\right)\right)^{2} + y_{i}^{2}}\end{equation}

If PEC are vertical, ellipse focus coordinates are $(0,-e'a')$ and $(0,e'a')$ and then
\begin{equation}r^{(1)}_{ij} = \sqrt{x_{i}^{2} + \left(y_{i} + \sqrt{\left(a + \frac{\sigma}{2}\right)^{2} - \left(b + \frac{\sigma}{2}\right)^{2}}\right)^{2}}\end{equation}
\begin{equation}r^{(2)}_{ij} = \sqrt{x_{i}^{2} + \left(y_{i} - \sqrt{\left(a + \frac{\sigma}{2}\right)^{2} - \left(b + \frac{\sigma}{2}\right)^{2}}\right)^{2}}\end{equation}\\

\end{document}